# Using SPICA* Space Telescope to characterize Exoplanets

## A White Paper for
## ESA's Exo-Planet Roadmap Advisory Team, July 2008

**ABSTRACT:**

We present the 3.5m SPICA* space telescope, a proposed Japanese-led JAXA-ESA mission scheduled for launch around 2017. The actively cooled (<5 K), single aperture telescope and monolithic mirror will operate from ~3.5 to ~210 μm and will provide superb sensitivity in the mid- and far-IR spectral domain (better than JWST at λ > 18 μm).  SPICA is one of the few space missions selected to go to the next stage of ESA's Cosmic Vision 2015-2025 selection process. In this White Paper we present the main specifications of the three instruments currently baselined for SPICA: a mid-infrared (MIR) coronagraph (~3.5 to ~27 μm) with photometric and spectral capabilities (R~200), a MIR wide-field camera and high resolution spectrometer (R~30,000), and a far-infrared (FIR ~30 to ~210 μm) imaging spectrometer - SAFARI - led by a European consortium. We discuss their capabilities in the context of MIR direct observations of exo-planets (EPs) and multiband photometry/high resolution spectroscopy observations of transiting exo-planets. We conclude that SPICA will be able to characterize the atmospheres of transiting exo-planets down to the super-Earth size previously detected by ground- or space-based observatories (e.g. radial velocity, super-WASP, COROT, KEPLER, TESS). It will also directly detect and characterize Jupiter/Neptune-size planets orbiting at larger separation from their parent star (>5-10 AU), by performing quantitative atmospheric spectroscopy and studying proto-planetary and debris disks. In addition to these scientific advances, SPICA will be a scientific and technological precursor for future, more ambitious, IR space missions for exo-planet direct detection as it will, for example, quantify the prevalence exo-zodiacal clouds in planetary systems and test coronagraphic techniques, cryogenic systems and lightweight, high quality telescopes.

**WRITTEN by**

J.R. Goicoechea[1], B. Swinyard[2], G. Tinetti[3], T. Nakagawa[4], K. Enya[4], M.Tamura[5],
M. Ferlet[2], K.G. Isaak[6] and M. Wyatt[7] on behalf of the *SPICA instrument teams*,
**and**
A.D. Aylward[3], M. Barlow[3], J.P. Beaulieu[8],  A. Boccaletti[9], J. Cernicharo[10], J. Cho[11],  R. Claudi[12], H. Jones[13], H. Lammer[14], A. Leger[15], J. Martín-Pintado[10], S. Miller[3], F. Najarro[10],  D. Pinfield[16], J. Schneider[9], F. Selsis[17], D.M. Stam[18], J. Tennyson[3], S. Viti[3] and  G. White[19].

[1]UCM, Madrid (**Spain**), [2]RAL (**UK**), [3]UCL (**UK**), [4]ISAS (**Japan**),[5]NAO (**Japan**), [6]University of Cardiff (**UK**), [7]University of Cambridge (**UK**), [8]IAP (**France**), [9]Observatoire de Paris-Meudon (**France**), [10]DAMIR, CSIC (**Spain**), [11]Queens Mary University of London (**UK**), [12]INAF (**Italy**), [13]University of Hertfordshire (**UK**), [14]Austrian Academy of Science (**Austria**), [15]IAS (**France**), [16]University of Hearts (**UK**), [17]University of Bordeaux (**France**), [18]SRON (**The Netherlands**)  and [19]Open University (**UK**).

*\* The Space Infrared Telescope for Cosmology and Astrophysics (SPICA)
   is a proposed  joint mission between JAXA (Japan) and ESA (Europe).*          - COVER PAGE -

# 1. Introduction

The study of exo-planets requires many different approaches across the full wavelength spectrum to both discover and characterise the newly discovered objects in order that we might fully understand the prevalence, formation and evolution of planetary systems. The mid infrared (MIR) region (from ~3 or 4 µm to ~30 µm) is especially important in the study of planetary atmospheres as it spans both the peak of thermal emission from the majority of exo-planets thus far discovered (200-1000 K) and is particularly rich in molecular features that can uniquely identify the composition of planetary atmospheres and trace the finger prints of primitive biological activity. In the coming decades many space and ground based facilities are planned that are designed to search for exo-planets on all scales from massive, young "hot Jupiters", through large rocky super-Earths down to the detection of exo-Earths within the "habitable" zone (the latter being, at present, a distant goal for a future interferometer or optical coronagraph in space). Few of the planned facilities, however, will have the ability to characterise the planetary atmospheres which they discover through the application of MIR spectroscopy. The Japanese led SPace Infrared telescope for Cosmology and Astrophysics will be realized within 10 years and has a suite of instruments that can be applied to the detection and characterization of exo-planets over the wavelength range from 5 to 210 µm – i.e. mid to far infrared. In this paper we briefly describe the SPICA mission and its complement of instruments and discuss the scientific advantages of using the MIR for exo-planet research. We highlight the opportunity that SPICA represents to the European exo-planetary community and strongly urge ESA to continue its participation in the mission.

# 2. The SPICA Mission and the focal plane instruments

SPICA will have the same size telescope as the ESA Herschel Space Observatory (3.5 m), but cooled to < 5 K thus removing its self emission and delivering an improvement in photometric sensitivity over Herschel of two orders of magnitude in the far infrared (FIR) range. This sensitivity, combined with a wide field of view and coverage of the full MIR/FIR waveband, will revolutionise our ability to determine the nature of the thousands of objects that Herschel, JWST, and SPICA, will discover in photometric surveys. The SPICA telescope will be monolithic, unlike the segmented JWST mirror, and will deliver diffraction limited performance at 5 µm with a clean point spread function (PSF). This will allow SPICA to have, in principle, the ability to provide MIR coronagraphy for exo-planet imaging with a high rejection ratio and a small inner working angle (IWA); we describe below how this will be achieved using a dedicated MIR coronagraph.

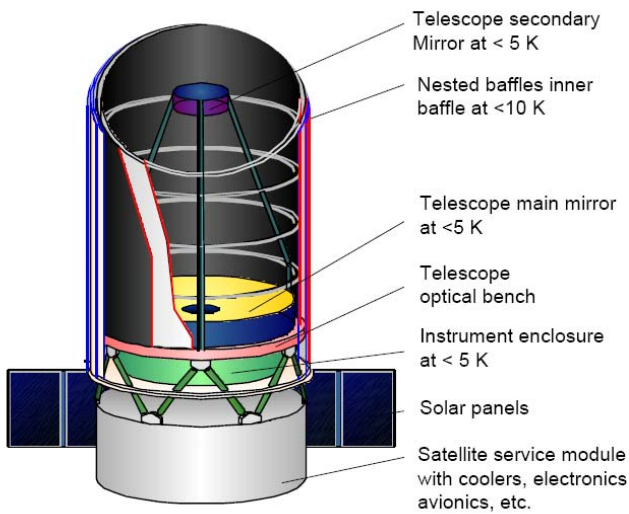

SPICA is planned for launch by JAXA on the H2A launcher in 2017 with a nominal five year mission lifetime orbiting at L2. The basic layout for the satellite is shown in figure 1. The telescope and instruments will be cooled using a combination of high reliability mechanical coolers and passive radiative cooling giving a long lifetime and low launch mass for the mission (2.6 tonnes). At least three focal plane instrument functions are proposed: an MIR instrument with broadband imaging and spectrographic capabilities to be developed by a Japanese/South Korean consortium; an MIR coronagraph to be developed by Japanese institutes and a (nationally funded) FIR imaging spectrometer to be developed primarily in Europe/Canada with possible contributions from Japan and the USA (NASA).

**Figure 1**: *Basic components of the SPICA satellite*

The SPICA mission was proposed as a possible joint mission to ESA as part of the Cosmic Vision 2015-2025 exercise and accepted for an assessment study in late 2007. This proposal called for ESA to provide the telescope, support for the ground segment and engineering support to a nationally provided FIR instrument. An industrial study is underway to assess the feasibility and cost of the telescope and ESA are looking at the provision of the Cebreros ground station for SPICA. A consortium of European national institutes is studying the design and provision of the FIR instrument (named SAFARI).

In Tables 1 and 2, we summarise the outline of the specifications of the three SPICA focal plane instruments. In practice it is planned to have two focal plane instruments: one for MIR and the other for the FIR. The coronagraph instrument will be implemented as an independent module to the cameras and spectrometer within the MIR instrument.

Table 1: *Specifications for the MIR camera and spectrometer*

| Module | | Wavelength coverage (µm) | FOV (arcsec) | Detector |
|---|---|---|---|---|
| Wide field camera | ch1 | 5-9 | 200x200 | Si:As 1k x 1k x 4 |
| | ch3 | 14-27 | | Si:As 1k x 1k |
| | ch2 | 8-15 | 200x200 | Si:As 1k x 1k |
| | ch4 | 20-38 | | Si:Sb 1k x 1k |
| High dispersion spectrometer R ~30000 | Short | 4-8 | 10x5 | Si:As 1k x 1k |
| | Long | 12-18 | 10x5 | Si:As 1k x 1k |
| Long wavelength spectrometer R~few 1000 | Short | 16-25 | 20x10 | Si:As 1k x 1k |
| | Long | 24-38 | 20x10 | Si:Sb 1k x 1k |
| Coronagraph R~200 | Short | 3-5 | 25x25 | InSb 1k x 1k |
| | Long | 5-20 | | Si:As 1k x 1k |

Table 2: *Specifications of the SAFARI Instrument*

| Parameter | Specification |
|---|---|
| Wavelength Coverage | At least 35 to 210 µm with a goal of 30 to 210 µm |
| Operational Modes | Photometric camera with at least three bands centred at approximately 50, 90 and 160 µm with R~3 to 5 |
| | Spectroscopy over the full wavelength range with at least R = 2000 at 100 µm and a goal of R = 10000 at 100 µm |
| Instantaneous wavelength coverage | As large as possible for spectroscopy |
| | At least three photometric bands simultaneously viewing the sky |
| Field of View | Minimum of 2x2 arcmin for camera mode |
| | Minimum of 1x1 arcmin for spectroscopy |
| Line sensitivity | Requirement <10 x$10^{-19}$ W m$^{-2}$ (5-σ 1 hour) with goal of ~1x$10^{-19}$ W m$^{-2}$ |
| Continuum sensitivity | Requirement <50 µJy (5-σ 1 hour) with goal of <20 µJy |

**The Coronagraph**

Given the importance of MIR coronagraphy to the study of exo-planets, we give here some more details of how the instrument might be implemented. It will provide both imaging with broad/narrow band filters and spectroscopy with R~200. A more detailed specification is given in table 2 and the optical chain and principles of operation described in figure 3. Note that the wavelength coverage is currently set as 5-27 µm with a possible extension to 3.5 µm. These limits are set by the Si:As array to be used as the primary detector. Even if the telescope can be made with a sufficiently high quality, a further extension below 3.5 µm is considered as a stretch goal as it will require a different detector technology (InSb). The specification is for the raw contrast only, and it is expected that PSF subtraction of the star will improve the contrast by a factor of ~10. The baseline design is to employ a binary shaped pupil mask designed to avoid the support structure diffraction spikes. The implementation of Phase Induced Amplitude Apodization (PIAA) is being developed to add even higher performance to the instrument. The IWA of the instrument using a binary mask on its own would be ~3.5λ/D; using the combination of a binary mask and PIAA can reduce this to ~2λ/D. To reduce the impact of the residual fixed (DC) wave front error from the telescope, a 32x32 pixel deformable mirror will be used in the optical train. Laboratory demonstrations of both of the binary mask and PIAA coronagraph have successfully achieved contrast better than $10^{-6}$ using a visible He-Ne laser. A prototype cryogenic MEMS type deformable mirror has also been demonstrated.

Table 3: *Detailed specification of the SPICA coronagraph instrument.*

| Parameter | Specification |
|---|---|
| Core wavelength (λ) | 5−27 micron (3.5-5 micron is optional) |
| Observation mode | Imaging, Spectroscopy |
| Coronagraphic method | binary shaped pupil mask or hybrid (binary shaped pupil + PIAA) |
| Inner working angle (IWA) | ~ 3.5 × λ/$D^{*}$ (binary shaped pupil mask mode) < 2 × λ/$D^{*}$ (hybrid mode) |

| Parameter | Specification |
|---|---|
| Throughput | ~30% (binary shaped pupil mask mode) <br> ~80% (hybrid mode) |
| Outer working angle (OWA) | ~ 30 × $\lambda/D$ (binary shaped pupil mask mode) <br> ~ 10 × $\lambda/D$ (hybrid mode) |
| Contrast between IWA and OWA | $\geq 10^6$ |
| Detector | 1k × 1k format Si:As array, 0.1''/pixel |
| Field of View | 1' × 1' |
| Spatial resolution | < IWA |
| Spectral resolution | ~ 200 |

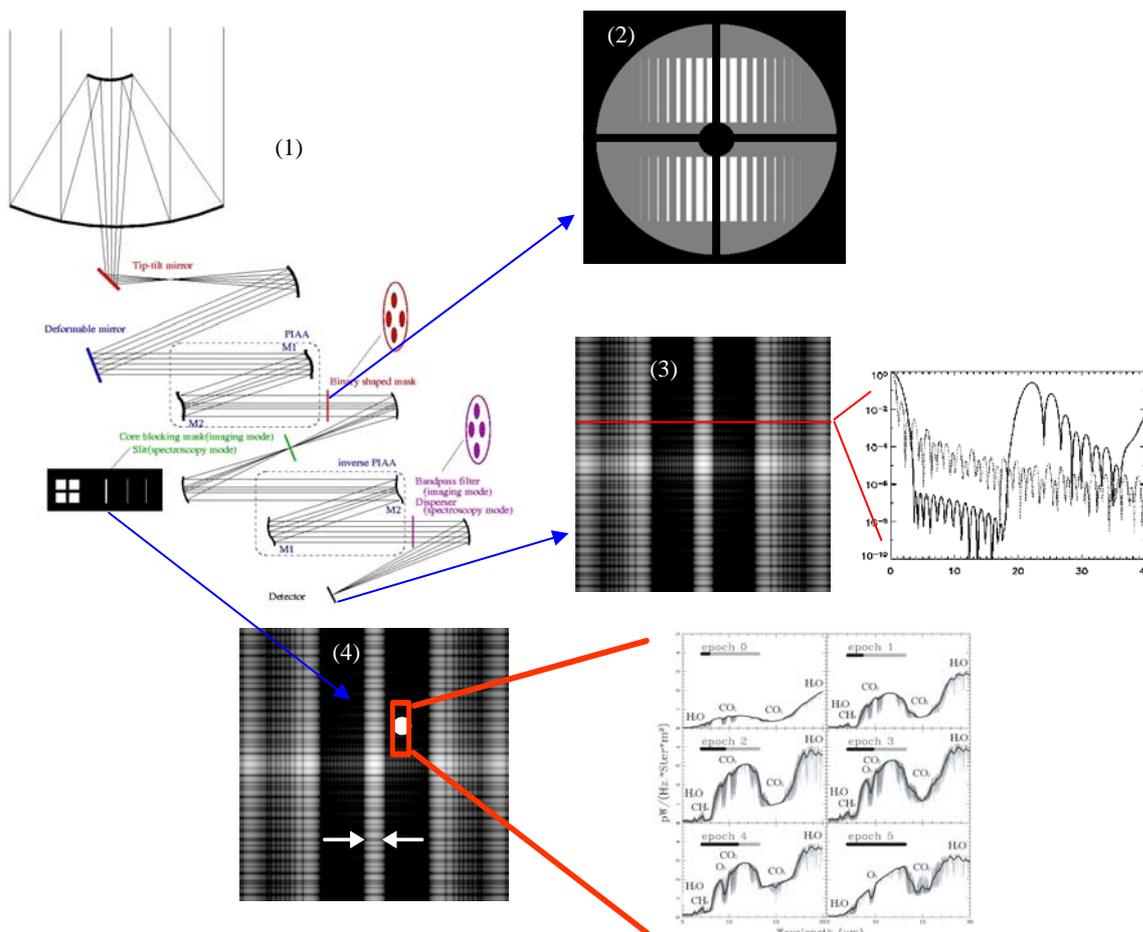

**Figure 2**: *Composite diagram showing the optical layout of the coronagraph (1); the shape of the binary mask placed at a pupil image within the optical train (2) (note the cross representing the direction of the diffraction spikes from the secondary supports); the image recorded at the detector (3) together with a slice though the image showing the suppression of the star's diffraction pattern with (solid line) and without (dotted line) the presence of the coronagraph. The final image (4) shows how the spectroscopy function will work by placing a slit over the planet image and selecting a grism/prism to disperse the spectrum – possible Earth like planet spectra from Kaltenegger, Traub, Jucks (2007).*

## 3. SPICA's Capabilities and Exo-Planet Research

The discover and exploration of extra-solar planetary systems is a young, promising and diverse field. Since the first discovery of an exo-planet (EP) by Mayor and Queloz in 1995, the number of detected EPs has increased to >300 objects distributed over >260 planetary systems. Our current understanding of the physical and chemical properties of these EPs, as well as their formation histories, is however very limited. Quantitative progress in the field requires challenging observations over a large wavelength range (from UV to FIR) with different techniques (e.g. direct imaging, transits etc) as well as new theoretical developments (from simulations of EP formation in proto-planetary disks to models of the exo-atmospheric chemistry). In the following we summarise our view of the main steps needed to detect and characterise EPs, and how the instruments on board SPICA will drive EP science, well in advance of other dedicated and more complex missions (e.g. TPFs).

## 3.1 What new steps are needed in the detection and characterisation of Exo-Planets

Exo-planetary systems may include a variety of planets even richer than the selection we have in our own Solar System: gas-giants, ice-giant planets as well as less massive icy/rocky planets (e.g., Guillon et al. 2007, Fortney et al. 2008). In addition, exo-zodiacal emission from debris associated with either a rocky zone of asteroids (warm) or a Kuiper-belt of comets (cold) has been inferred in several systems. Most EPs detected to date have been found through radial velocity searches, and are massive gas planets ("hot jupiters", $T_{eff}$ ~1000 K) which orbit very close (<< 1 AU) to Solar-type stars, and thus have short periods (P<10 days). Planets with larger orbital axes – e.g. comparable to that of Jupiter or Saturn, could be detected more efficiently in the near future through ground-based direct detection capabilities (VLT-SPHERE, GEMINI-GPI) or astrometry methods (e.g. with GAIA and SIM). However, the characterisation of these objects requires not just the detection of their presence, but also the detailed evaluation of their atmospheric composition.

The dramatic measurements made with the Spitzer and Hubble Space Telescopes over the past 5 years have redefined the field of EP characterisation (e.g., Barman 2007, Beaulieu et al. 2008, Charbonneau et al. 2002, 2005, Deming et al. 2005, 2006, 2007, Marley et al. 2007, Demory et al. 2007, Gillon et al. 2007, Grillmair et al. 2007, Harrington et al. 2006, 2007, Knutson et al. 2007, Richardson et al. 2006, 2007, Machalek et al. al 2008, Swain et al. 2008a,b; Tinetti et al., 2007). Collectively, this work has conclusively established that the detailed characterisation of exo-planet atmospheres is feasible. Today, due to the extraordinary and unforeseen success of the Spitzer and Hubble, we can discuss the observational signatures of EP atmospheres including weather, vertical and longitudinal temperature profiles, molecular abundances (including prebiotic species or biomarkers; e.g. Selsis et al 2002), dayside to nightside atmospheric chemistry changes, and the role of photochemistry. Spitzer and Hubble measurements do not spatially resolve the planets; rather, information about the planet is extracted from precise measurements in the "light curve" or changes in the spectrophotometric intensity arising from the planet during its orbit. Further, observations of different portions of the light curve have demonstrated the ability to localize molecular abundances to specific regions of the planetary atmosphere and explore, for example, the differences of dayside and nightside atmospheric chemistry.

Spitzer has also measured EP photometric eclipses out to 24 µm demonstrating that the light-curve is simpler ("box-like") than in the visible domain due to the absence of stellar limb-darkening effects. This allows a robust determination of the EP radius as a function of wavelength and provides further strong constraints to the atmospheric properties. Future space missions operating in the same waveband will need to achieve a photometric precision better than the ~0.02% currently reachable with the Spitzer instruments if the atmospheric features of hot-Jupiters and Neptunes are to be detected. Characterisation of the atmosphere of an Earth-size planet with a hot, volatile-rich atmosphere in transit will require a photometric precision of a factor of ~ 10 higher. Spitzer has demonstrated that the low resolution EP spectra can be extracted even without imaging the planet directly, by subtracting the stellar spectrum during the secondary eclipse from that observed outside the eclipse. In particular, transit observations with the IRS spectrometer have been used to extract the absolute intrinsic spectrum of HD209458b planet (d~47 pc, G0 primary star) - with the resultant spectrum in physical units (e.g., Jy) as opposed to the typical relative contrast measurements. Such observations require a complicated calibration scheme able to reach a <0.1% of absolute calibration accuracy. This is achieved by a variety of methods, including: removal of telescope pointing errors and drifts, zodiacal light subtraction and "star leakage" contribution (Swain et al. 2008c). The low spectral resolution used (R = 60-120) and the poor S/N do not, however, always allow a unique interpretation of the atmospheric features (Swain et al., 2008b).

The observational techniques used for transiting systems can also be extended to non-transiting systems, so it is valuable to consider a generalization of the transit technique, namely EP characterization in "combined light". Without a transit, the planet radius cannot be measured directly, which is a significant limitation. Nevertheless, much can be learned, for example from observing fluctuations in IR intensity that are phased to the planet's known radial velocity orbit (e.g., Harrington et al. 2006).

Future observations are expected to directly address our understanding of the formation and evolution of EP systems and to engage in the search for spectral signatures from atmospheric biomarkers and non-equilibrium chemistry. To make progress in the field beyond present day "indirect" detections and low resolution MIR spectra requires: *(i)* "direct imaging" methods: to separate the planet's light photons from those of the host star and *(ii)* higher sensitivity and higher spectral resolution spectroscopic studies: to detect the main chemical constituents of EP atmospheres. The latter could be either by *"direct spectroscopy"* with coronagraphs, *"transmission spectroscopy"* during primary eclipses, *"occultation spectroscopy"* during secondary transits or *"combined light"* spectroscopy of both star and planet. The huge flux contrast between the star and planet (from ~$10^{-9}$ in the visible to ~$10^{-6}$ in the mid-IR) and the small planet-to-star angular separation (e.g. the distance from Saturn to the Sun seen from 10 pc is ~1'') makes these goals extremely challenging from a technical point of view.

Following on from the pioneering observations carried out using the Hubble and Spitzer Space Telescopes, the "immediate" successor missions (e.g., JWST and SPICA) should be able to observe and characterise planets down to the Earth-size orbiting close in or within the habitable zone of M-stars. By adding a coronagraph, it will be possible to study cooler gas-giants at intermediate orbital distances (> 5-10 AU) around "Sun-like" stars and cooler K, M dwarfs - the most numerous stars in the galaxy. Further into the future, EP-specialised missions will be able to detect directly and characterise less massive and cooler Earth-like rocky planets (<10 $M_{EARTH}$) in the "habitable zone" of G and K stars. These missions will focus on the detection of bio-signatures, and should be able to probe surface properties including the presence of liquid water (e.g. by combining information of radiation intensity and polarization in the visible spectral range).

In addition, SPICA will be extremely effective at characterising the debris disks of nearby stars. The dust detected from such disks is indicative of the presence of analogous asteroid and Kuiper belts (Wyatt 2008), or of regions where the formation of Earth-like or Pluto-like planets is ongoing (Kenyon & Bromley 2008). Thus, in addition to the need to characterise this dust emission so that its impact on the direct detection of planets can be mitigated (Beichman et al. 2006), these disks are of fundamental interest for their description of the current state of planet formation and of planetary system architecture (Wyatt 2008). Furthermore, the nearest debris disks can be imaged, and it is already possible to indirectly infer from observations of clumpy /warped /asymmetric dust structures that planets as small as Neptune are gravitationally perturbing the planetesimals and dust (Wyatt et al. 1999, 2003; Kalas et al. 2005).

## 3.2 How can SPICA push the boundaries

### 3.2.1 The Mid-IR coronagraph

Direct detections with coronagraphs (also foreseen for JWST, see below) provide a huge advantage compared to transit observations as one directly extracts the photons coming from the planet. In spite of the different telescope sizes of SPICA (D~3.5m) and JWST (D~6.5m), the projected coronagraph on board SPICA has several advantages over the JWST coronagraphs. In this comparison we shall restrict ourselves to JWST-MIRI, which roughly covers the same wavelength domain as SPICA's MIR instruments. First, the monolithic mirror of SPICA will be better optimised for coronagraphy than the segmented mirror of JWST due to its much simpler and clean PSF. The segmented geometry also requires complex Lyot stops for the suppression of the light diffracted by each mirror segment, reducing the throughput and requiring a high degree of alignment stability. Secondly, the SPICA telescope itself will be actively cooled down to 4.5 K (compared to passive cooling down to 45 K for JWST): it is, therefore, optimised for mid- and far-IR astronomy. From the scientific point of view, one of the major differences between the two facilities is the possibility of undertaking coronagraphic spectroscopy with SPICA (using a grism/prism providing R~200) in addition to coronagraphic imaging. This spectral capability in a wavelength domain rich in chemical signatures differentiates the unique science possibilities of SPICA (Abe et al. 2007) compared to JWST/MIRI, which will only have quadrant phase and Lyot photometric coronagraphs at ~10.7, ~11.4, ~15.5 and 23 μm.

Direct imaging of rocky earth-like planets within the "habitable zone" requires subarsec angular resolution in order to remove the bulk of the signal originating from the central star (e.g. TPF of Darwin type missions). However, the "core accretion" model for the formation of extrasolar giant planets (EGP; Wetherill & Steward 1989) suggests that massive planets (>10 $M_{Earth}$) accrete a gaseous envelope from the surrounding planetary nebula. One thus expects that EGPs have thick atmospheres (controlling the incident stellar radiation and regulating the planet's thermal balance) of roughly nebular composition surrounding a denser core.

SPICA will add greatly to the field of exo-planet research (and much earlier) by directly imaging and, for the first time, recording the MIR spectra of EGPs, constraining their temperature and accessing their atmospheric chemistry. We estimate that a binary mask type coronagraph (simpler approach) will achieve a contrast of $10^{-6}$ at the equivalent to ~9 AU (~Saturn's orbit) at ~5 μm at 10 pc. At this wavelength we probe the younger end of the planet age range with ~30 target stars within 10 pc. Developments in coronagraphic performance through hybrid techniques (see above) will reduce this to ~5 AU at 10 pc (Jupiter's orbit) and, if the short wavelength extension is implemented, SPICA may image EGPs as close as 4 AU at 10 pc or see Saturn type planets out to 30 pc opening up the possibility of observing 100's of candidate systems. At the longest wavelengths (~27 μm) the IWA increases to ~50 AU at 10 pc. A complete MIR spectrum at R~200 could therefore be taken for EPs at moderate angular distances from the host star (>1''). A few of such planets are known presently, with many more likely to be discovered in the decade between now and SPICA's launch.

Hence, the targets for SPICA's coronagraph will be outer self-luminous, hot EGPs with ages between 1-5 Gyr. The high stellar suppression with SPICA's mid-IR coronagraph makes it perfectly suited also to debris disk imaging. It is particularly important that this set-up can image both planets and debris simultaneously so that the interaction between these components both during and after planet formation can be explored.

The most significant atmospheric features expected in the EP spectra of SPICA's coronagraph targets can be summarized as follows; *(i)* The ~4-5 µm "emission bump" due to an opacity window in EGPs and cooler objects with temperatures between 100 and 1000 K. *(ii)* Molecular vibration bands of $H_2O$ (~6-8 µm), $CH_4$ (~7.7 µm) , $O_3$ (~9.6 µm), silicate clouds (~10 µm) , $NH_3$ (~10.7 µm), $CO_2$ (~15 µm) and many other lower abundance species such as hydrocarbons and more exotic nitrogen/sulphur-bearing molecules. If detected, the relative abundance of all these species could be compared among different EPs and with Solar System planets and bodies. *(iii)* $He-H_2$ and $H_2-H_2$ collision induced absorption band features around ~17 µm as tracer of the He/H relative abundance. Note that it is well established that the He abundance spans large variation in the gaseous planets of the Solar System. *(vi)* Features from deuterated molecular species to distinguish cool brown dwarfs from "real" EP (e.g., $CH_3D$ at ~8.6 µm) and non-equilibrium species (e.g. $PH_3$ at ~8.9 and ~10.1 µm). *(v)* Finally, information on the EP weather/dynamics or the day/night-side spectra could potentially be obtained from precise monitoring of the time variation of these spectra.

SPICA has been designed as true observatory with an extensive range of astronomical targets and science goals and will not be optimized for EP searches. However, its high sensitivity and high-contrast coronagraph will allow the search for new EGPs in a few stellar systems with appropriate physical conditions for the formation of planets (e.g. luminosity, stellar gravity, metallicity and distance). Optimising coronagraphic EP searches is not an obvious task and several survey strategies could be adopted (e.g. Agol 2007). The largest uncertainty in EP detection efficiency is due to exo-zodiacal dust. As we will see in the next sections, the two other instruments on SPICA could be used to detect and characterise any zodiacal background in a statistical sample of stars.

### 3.2.2 The Mid-IR camera and high-resolution spectrometer

The MIR instrument on SPICA discussed in the previous section has a wide-field camera, a long-slit grating spectrometer with an Integral Field Unit (IFU) and an immersion grating which will be used to achieve a spectral resolution as high as R~30,000 (~10 km/s), an order of magnitude larger than the resolution provided by JWST. In spite of the smaller size of SPICA telescope its much lower operating temperature will make the SPICA instrument virtually as sensitive as JWST-MIRI, i.e. around ~1 µJy in photometric mode (5σ-1h). This means that SPICA can go deeper than $M_V$=20 and detect M dwarfs and other faint objects. Spitzer/IRS's studies of the HD 209458b transit have shown that the intrinsic flux of this archetypical hot Jupiter is ~200 µJy at 15 µm (see figure 5); well above the sensitivity expectations of SPICA. Therefore, SPICA's MIR instrument could be used to perform *(i)* transit photometry (e.g., to constrain the planet radius as a function of wavelength) and *(ii)* perform "transmission" and "occultation" spectroscopy (to extract the thermal continuum and spectral features) in a large statistical sample of EPs. In particular, we estimate that the ~30 µm thermal continuum of EPs similar to HD 209458b could be observed in transit observations of systems as far as ~150 pc (~250 stars) as the contrast requirement is modest: ~0.5% at the shortest wavelengths. Another step forward in transit studies will be the observation of cooler and/or older EPs (real Jupiter-like planets). Note that the thermal emission of a planet with R = $1R_J$ and $T_{eff}$ ~120 K peaks at ~25 µm (where SPICA starts to become more sensitive than JWST). The superb sensitivity of SPICA instrumentation means that this kind of cool planet around stars at d < 5 pc (and with low exo-zodiacal backgrounds) could be characterised during their eclipses if a contrast of ~0.01% can be achieved.

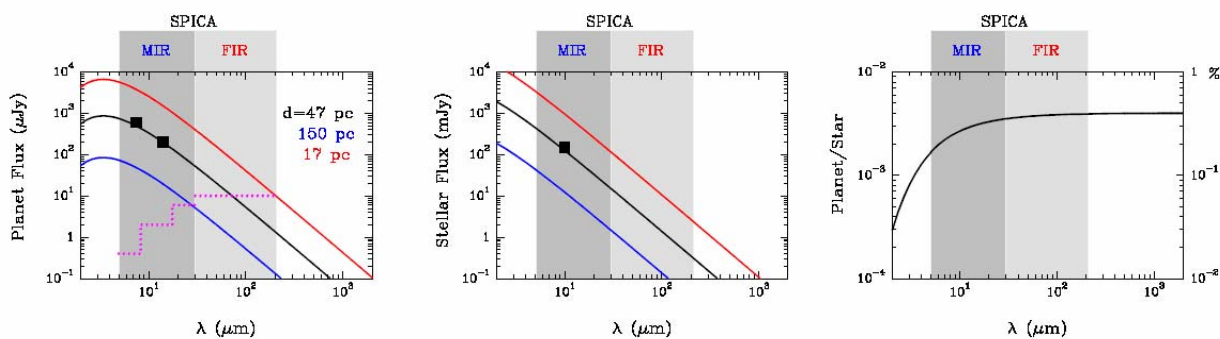

**Figure 3** - *Left: Fit to HD 209458b hot Jupiter ($T_{eff}$~1000 K) mid-IR fluxes obtained from Spitzer/IRS observations (Swain et al. 2008c) during its transit around a G0 star (d~47 pc, $T_{eff}$ ~6000 K) and estimations for different distances (blue and red curves). The magenta dashed lines show the estimated photometric sensitivity of SPICA with GOAL detectors (5σ-1hr). Middle: Expected flux from the host star. Right: Resulting Planet-to-star contrast ratio as a function of wavelength. The shaded regions represent the 2 main wavelength domains of SPICA instruments: "MIR" for the coronagraph and mid-IR instruments and "FIR" for the SAFARI instrument.*

Transmission spectroscopy with the mid-IR high-resolution spectrometer will probe the low-pressure layers of EP atmospheres where non-equilibrium chemical conditions are more likely to occur. The first optical transmission spectrum taken from the space used a medium resolution spectrometer (R~5500; Charbonneau et al. 2002). More recently, high resolution (R~60,000) ground-based optical transmission spectra have been used to resolve the EP line features (Redfield et al. 2008). High-S/N, high-resolution MIR spectra of transiting EPs will provide a unique opportunity to determine the properties of EP atmospheres and exospheres through detailed analyses of absorption band profiles and careful comparison with atmospheric models. Such observations will resolve spectrally the band profiles of several key molecular vibration bands expected in EP atmospheres towards bright stars. The improved spectral resolution will also allow the detection of weaker bands from less abundant atmospheric species (spectrally diluted at low resolution). Target bands in each of the baselined channels of the instrument include: $H_2O$ (most of IR domain) and $CH_4$ (~7.7 µm) in channel 1 (3.5-9 µm); $O_3$ (~9.6 µm), silicate clouds (~10 µm), $NH_3$ (~10.3 and ~10.7 µm), benzene (~14.8 µm) and acetylene (~13.7 µm) in channel 2 (8-15 µm), $CO_2$ (~15 µm – a strong feature in the Earth's atmosphere), $HC_3N$ at ~15 µm, $C_4H_2$ (~16 µm), $H_2O$ (~18 µm) in channel 3 (14-28 µm). The MIR instrument will also have access to the polycyclic aromatic hydrocarbon (PAHs) band features (at ~6, ~7, ~8, and ~11 µm) which are excellent tracers of UV-radiation fields in protoplanetary disks. Were the spectral coverage to be extended down to 3.5 µm then $H_3^+$ (~4 µm) would also be accessible, and with it the key to understanding the escape processes in the upper atmosphere and the atmospheric cooling mechanisms (e.g., Koskinen et al., 2007). In addition, high-spectral resolution may allow the detection of narrow molecular lines in the MIR: ro-vibrational lines of $H_2$, high-excitation pure rotational lines of light species such as HD, $H_2O$ or OH as well as atomic fine-structure lines from metals in different ionization degrees (e.g., Ne, Si, Fe, Ar, S, Ni, P, etc).

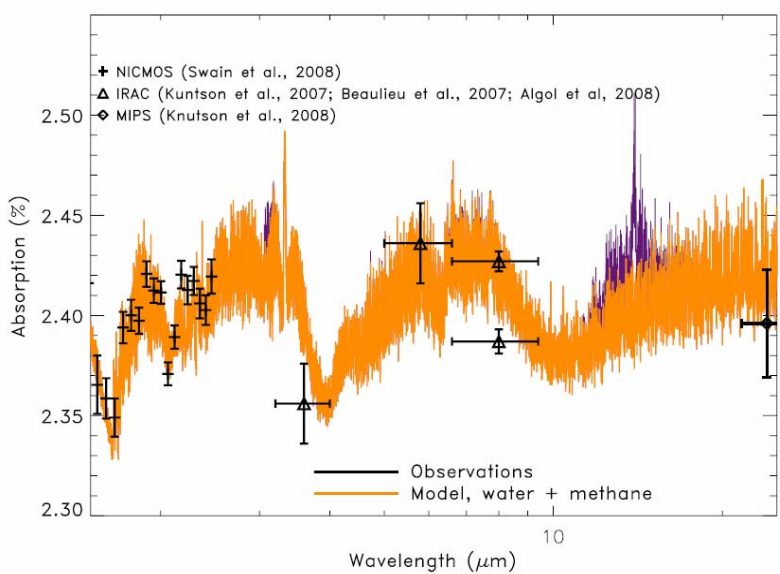

**Figure 4:** *A comparison of Hubble and Spitzer primary transit observations of HD 189733b hot Jupiter (black crosses, triangles and diamonds) and high-resolution simulations of the exoplanet atmosphere. Models include bands of water ($H_2O$), methane ($CH_4$) and hydrogen cyanide (HCN, in violet). The vertical axis corresponds to the expected absorption, the percentage of the star covered by the EP transit (Tinetti et al.). The horizontal axes corresponds to the wavelength in microns. The MIR instrument on board SPICA will be able to detect such features at high-spectral resolution (R~30,000) over the ~4 to 18 µm window on a large sample of transiting exoplanets.*

A high priority for SPICA using the combined-light method would be observations of habitable zone planets for *(i)* detection of pre-biotic molecules and *(ii)* quantification of the role of photochemistry and thermochemistry in atmospheric carbon chemistry. The MIR instrument will thus be able to perform multiband transit photometry of hot and warm EPs down to the super-Earth size, previously detected by COROT, KEPLER and TESS, or by radial velocity, as well as high-resolution transit spectroscopy for the brighter objects.

### 3.2.3 The European Far-IR instrument (SAFARI)

SAFARI will cover the FIR window that extends from ~30 µm (the upper cut-off of the MIR instruments) to ~210 µm (just long ward of the [NII] 206 fine structure line) with a field-of-view of 2'x2'. Assuming diffraction limited performance, SAFARI will provide angular resolutions from ~2'' to 15'' (20 to 150 AU at 10 pc) at wavelengths not covered by JWST and, as shown in figure 6, at more than 2 orders of magnitude higher sensitivity than Herschel/PACS. This huge increase in sensitivity could potentially open EP research to wavelengths completely blocked by the Earth's atmosphere, but representing the emission peak of many cool bodies (gas-giant planets, asteroids and so on). SAFARI will have its major strength in measuring excess radiation from dusty proto-planetary disks in hundreds of stars at almost all galactic distances. It will also perform medium spectral resolution observations (R~2,000) over a spectral range also rich in dust features, water vapour rotational lines (temperatures below ~500 K), atomic oxygen fine structure lines at ~63 µm and the solid state water-ice features at ~44 and ~62 µm.

Water ice is the major ingredient of the core of gas giant planets comets and smaller objects and its detection in proto planetary systems and, in the nearest objects, mapping will provide the observational evidence needed to constrain models of planetary formation and evolution.

The range of temperatures (~50-500 K) probed by SAFARI continuum and spectral diagnostics, combined with its projected instrument sensitivity, ~10 µJy (5σ-1hr), will provide capabilities that complement MIR coronagraph studies of EGPs. First, SAFARI will be able to characterise the exozodiacal dust component of several thousand nearby stars. Determining the exo-zodiacal background levels from observations of a large sample of stars will be key to prioritising Earth-like candidates for searches with longer-term TPF-type missions (due to increased photon noise and potential confusion with zodiacal structures). With two orders of magnitude in sensitivity over Herschel, SPICA would be able to survey stars to the level of dust mass that is limited by calibration accuracy (i.e., 0.01 lunar mass for 90K grains around Sun-like stars; Wyatt 2008) out to 10 times greater distance (e.g., to 180 pc as opposed to 18 pc, for Sun-like stars), resulting in 1000 times as many detections (e.g., of order $10^5$ rather than $10^2$ Sun-like stars can be surveyed with SPICA to this limit). Notably this increase in distance encompasses most of the nearby star forming regions meaning that SPICA will be particularly adept at studying both protoplanetary disks and the brief epoch at ~10Myr at which the transition from protoplanetary to debris disk occurs, a transition which is as yet poorly understood but which is of prime importance due to its curtailment of (and potentially direct link with) planet formation. Access to such sources is also important because giant planets are brightest at the youngest ages.

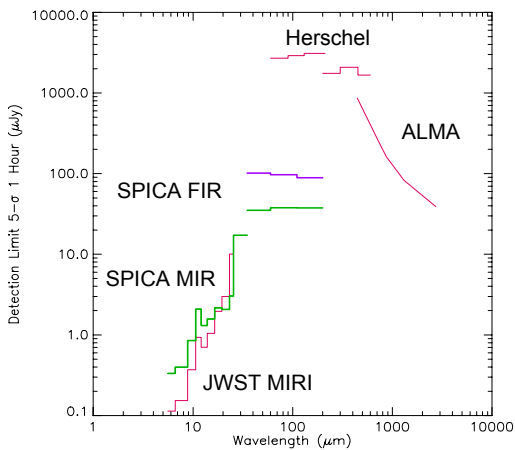 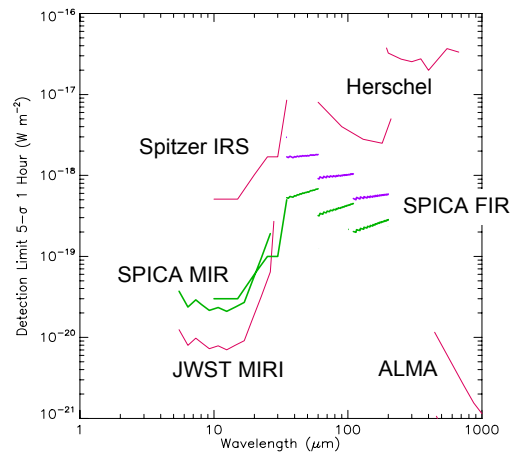

**Figure 5** - *Predicted photometric performance of SPICA (green [goal] and purple [requirement]) compared to predecessor and complementary facilities (red) given as point source sensitivities in µJy for 5-σ in 1 hour over the bands shown indicatively as horizontal lines. Note the 2 orders of magnitude increase in FIR photometric sensitivity compared to Herschel that will be achieved using goal sensitivity detectors on SPICA. The figures here are raw sensitivity with no allowance for confusion.*

**Figure 6** - *Predicted spectroscopic performance of SPICA (green [goal] and purple [requirement]) compared to predecessor and complementary facilities (red) given as single unresolved line sensitivity for a point source in W m$^{-2}$ for 5-σ in 1 hour. For ALMA 100 km/s resolution is assumed as this is appropriate for extra-galactic sources. The SPICA MIR sensitivities are scaled by telescope area from the JWST and Spitzer IRS values respectively.*

Secondly, with very stable detectors and efficient, high cadence and high S/N observations, SAFARI could also be used to perform transit photometry and, on the brightest candidates, spectroscopy for the first time in the FIR domain. FIR light curves will allow the characterisation of EP radii as a function of wavelength, whilst occultation spectroscopy, during secondary eclipses, may reveal the same bright rotational water emission lines detected by ISO in the atmospheres of Jupiter, Saturn, Titan, Uranus and Neptune (Feuchtgruber et al. 1999). Note that with SAFARI's sensitivities one could potentially extract the ~200 µm thermal continuum of EPs like HD 209458b at distances out to <20 pc (see Figure 3 for more detailed predictions). In addition, and also from transit events, it may be possible to extract the thermal emission of nearby massive-enough cool planets (100-200 K) as a detectable contrast only occurs at wavelengths longer than 30 µm.

In summary, it is likely that all the SPICA instruments will be able to perform follow-up observations of ground-based, COROT and KEPLER detected EGPs, making coronagraph searches of young hot Jupiters, performing quantitative atmospheric spectroscopy and studying proto-planetary and debris disks. SPICA will contribute to the study EPs in different phases of their evolution, revealing the different formation scenarios (e.g., rapid collapse through gravitational instabilities versus slower formation due to core accretion mechanisms) as well as the underlying chemistry (e.g., transport of water and volatile species into the inner planetary system, etc.).

# 4. Status of the Mission Study

Japan

The SPICA mission definition is relatively mature, having been developed and reviewed within Japan on several occasions in the last four years. SPICA is on the current Japanese space science roadmap and in February 2008, the mission formally passed the Japanese Mission Definition review and was approved by the Japanese Space Science Steering Committee in March 2008. The mission has now been to the JAXA Project Preparation review and has been approved to move into "Pre-Project" status with a JAXA project team starting a Mission Concept Study lasting from now until to August 2009. The scope of the Mission Concept Study is similar to a comprehensive Phase-A study. This would then lead to the Phase-B study concluding in late 2010 and commencement of the spacecraft Phase C/D in 2011. The SPICA mission is therefore up and running in Japan and in the medium term there will be a Systems Requirement Review (SRR) in mid 2009 to define the spacecraft resource envelope.

ESA

The internal ESA assessment of the SPICA telescope assembly and the ground segment support have been completed with the next step being the industrial assessment of the SPICA telescope assembly via ESA ITT – this is expected to start in the near future. The ESA internal assessment of the telescope concluded that the telescope could be procured entirely within Europe, although some development of a flight re-focussing mechanism was required. The specification of the telescope quality required to match the coronagraph requirements has been identified as a possible risk in terms of expense and the length of time required for polishing the mirror to a suitably high quality figure. Much of the "mid frequency" errors in the telescope figure can be compensated for using the 32x32 element deformable mirror discussed in section 2. If the development of this item proves problematic the specification of the telescope and coronagraph will need to be revisited to optimise the coronagraphic performance of SPICA at a systems level.

The ESA assessment of the SAFARI instrument was initiated by the drafting of an Instrument Definition Document based on input from the consortium. This was used as the basis of a Concurrent Design Facility (CDF) exercise with staff from ESTEC and the consortium involved to progress the instrument design and highlight areas where further technical development is required. The CDF exercise identified the detectors, mechanisms and cooling chain as requiring development but otherwise the instrument design has no significant issues.

In the longer term, ESA currently plan to down select from the current set of six candidate M-Class missions to two in the fourth quarter of 2009. These two missions then go into an Industrial definition phase leading up to the forth quarter of 2011 when a decision will be made to implement one or both of the M-Class missions. If SPICA is successful in this Cosmic Vision selection process, the timetable for the subsequent development phases for the ESA contributions and the SAFARI instrument are largely compatible with the overall JAXA SPICA schedule.

# 5. Summary

Spitzer and Hubble results demonstrate that we have, today, the flight-proven technology and the models required to study the prebiotic chemistry of EP atmospheres in "habitable zones". This is a completely unexpected yet extraordinary accomplishment. In a few years from now, SPICA could provide exciting further steps towards MIR direct detection of EP, and the characterization of their physical conditions, composition and atmospheric chemistry. In particular we have shown that:

- SPICA will be equipped with a MIR coronagraph (~3.5-27 µm) providing both photometric and medium-resolution (R~200) spectroscopy. This spectral ability in a wavelength domain rich in chemical signatures differentiates its unique science possibilities compared to those of JWST coronagraphs. This will allow SPICA to directly image and, for the first time, record intrinsic MIR spectra of outer EPs (>5-10 AU). Space "direct detections" have not been achieved yet.

- Despite its modest monolithic mirror size (D~3.5m), the SPICA telescope will be actively cooled down to ~4.5 K (compared to ~45 K through passive cooling for JWST), and thus will reach similar sensitivities to JWST/MIRI (around ~1 µJy) but with an increased wavelength range, wider field of view and increased observing efficiency due to the use of multiple detector arrays. SPICA's MIR high-resolution spectrometer will be able to achieve R~30,000, and thus to resolve spectrally key MIR bands of $H_2O$, $CH_4$, $O_3$, $NH_3$, $CO_2$, etc.

- In addition, SPICA will cover the longer FIR wavelengths (30 to ~210 µm), where it is orders of magnitude more sensitive than Herschel, thus opening the EP research to wavelengths completely blocked by the Earth's atmosphere, but at the emission peak of exo-zodiacal dust and cool bodies.

- The EGPs that will be studied with SPICA are "stepping stones" to detecting and characterizing earth-like planets through developing new techniques that will be used later for specialised searches of Earth-like planets using future planet finding facilities.

Building on the successes of Spitzer and Hubble in EP research, and given the improved scientific capabilities of SPICA, we strongly advocate that ESA continues its participation in the JAXA-led SPICA space mission beyond the present assessment period. As we have shown here, the European EP community would then have access to observing time possibilities to perform unique combined-light characterisation and direct imaging and spectroscopy of extra-Solar planets spanning a large variety of atmospheric properties, host star types and orbital distances.

## 6. Bibliography and references